\documentclass[twocolumn,showpacs,preprintnumbers,amsmath,aps]{revtex4}
\usepackage{graphicx}
\usepackage{dcolumn}
\usepackage{bm}
\usepackage{color}
\begin{document}
%
\def\eq#1{(\ref{#1})}
\def\fig#1{\ref{#1}}
\def\tab#1{\ref{#1}}
%
\preprint{{\bf sent to:} \textcolor{red}{{\bf Solid State Sciences}}}
\title{
---------------------------------------------------------------------------------------------------------------\\
Multigap superconducting state in molecular metallic hydrogen}
\author{R. Szcz{\c{e}}{\'s}niak, E.A. Drzazga}
\affiliation{Institute of Physics, Cz{\c{e}}stochowa University of Technology, Al. Armii Krajowej 19, 42-200 Cz{\c{e}}stochowa, Poland}
\email{edrzazga@wip.pcz.pl}
\date{\today}
\begin{abstract}
  The thermodynamic parameters of the superconducting state, that gets induced in the metallic molecular hydrogen under the influence of the pressure at $414$ GPa ($T_{C}=84$ K), have been determined. The calculations have been conducted in the framework of the three-band Eliashberg equations. 
  The order parameters ($\Delta^{\alpha}$) and the wave function renormalization factors ($Z^{\alpha}$) have been derived; the symbol $\alpha$ denotes the band index: $\alpha\in\left\{a,b,c\right\}$. It has been stated that the dimensionless ratios $2\Delta^{\alpha}\left(0\right)/k_{B}T_{C}$ are equal to: $5.55$, $3.96$ and $3.53$, respectively. Next, the total
normalized function of the density of states, the free energy, the thermodynamic critical field and the specific heat have been determined. The obtained results differ significantly from the results achieved in the framework of the one-band Eliashberg model for the comparable value of the pressure ($p=428$ GPa). 
  In the last step, the values of the wave function renormalization factors have been estimated for $T_{C}$. It allowed us to calculate the maximum value of the electron effective mass in a given band [($\left[m^{\star}_{e}\right]_{\alpha}$). The following results have been achieved: $\left[m^{\star}_{e}\right]_{a}=2.99 \left[m_{e}\right]_{a}$, $\left[m^{\star}_{e}\right]_{b}=2.10 \left[m_{e}\right]_{b}$ and $\left[m^{\star}_{e}\right]_{c}=1.94 \left[m_{e}\right]_{c}$, where $\left[m_{e}\right]_{\alpha}$ stands for the electron band mass. 
\end{abstract}
\maketitle

\vspace*{1cm}

Hydrogen is an element in which the superconducting state with a very high value of the critical temperature ($T_{C}$)
should be induced at the high pressure \cite{Ashcroft}. This claim is justified on the basis of the BCS expression \cite{BCS1}, \cite{BCS2}: 
\begin{equation}
\label{r1}
k_{B}T_{C}=1.13\omega_{D}\exp\left(-\frac{1}{\lambda-\mu^{\star}}\right), 
\end{equation}
where $k_{B}$ is the Boltzmann constant, $\omega_{D}$ is the Debye frequency, the symbol $\lambda$ denotes the electron-phonon coupling constant, and $\mu^{\star}$ represents the Coulomb pseudopotential. 

In the case of hydrogen, the Debye frequency reaches high values due to a very small mass of the atomic nuclei forming the crystal lattice (single protons); the electron-phonon coupling is strong due to the lack of strong internal electron shells. On the other hand, the depairing electron correlations modeled by the Coulomb pseudopotential are not high enough to significantly lower the critical temperature \cite{Maksimov}, \cite{R1}, \cite{R2}, \cite{R3}.

The studies on the thermodynamic properties of the superconducting state in hydrogen have lasted for many years. In particular, the {\it ab initio} calculations suggest that hydrogen becomes a subject of the metallization for the pressure at about $400$ GPa \cite{Stadele}. In the range of the pressure from $400$ GPa to $500$ GPa the superconducting state, characterized by a high value of the critical temperature ($242$ K for $p=450$ GPa) \cite{Cudazzo}, should be induced in the molecular phase of the metallic hydrogen. Additionally, it should be noted that the superconducting state in the molecular phase of the metallic hydrogen can be strongly anisotropic ($p=414$ GPa) \cite{Cudazzo}. In this case, an accurate description of the thermodynamic properties is possible only in the framework of either anisotropic model or multiband model \cite{Choi}, \cite{Nicol}, \cite{Dolgov}.

Above the pressure of $500$ GPa, the metallic phase of the molecular hydrogen becomes dissociated \cite{Stadele}, \cite{Johnson}, \cite{Pickard}. The results of the calculations performed for the metallic atomic hydrogen just above the
dissociation pressure ($p=539$ GPa) suggest a high critical temperature value ($T_{C}=357$ K); other thermodynamic parameters differ significantly from the values predicted by the BCS theory \cite{R4}. 

For extremely high pressures (above $1$ TPa), the superconducting state properties have been analyzed in the works \cite{Maksimov}, \cite{R5}, \cite{McMahon}. It has been stated that for $p=2$ TPa, the critical temperature may reach a record level of an order $600-700$ K. It is worth to underline that for $p=2$ TPa, the ratio of the energy gap to the critical temperature ($R_{\Delta}$) changes from $6.17$ to $6.63$ according to the assumed value of the the Coulomb pseudopotential. Similarly high values of the parameter $R_{\Delta}$ can be found only in the family of the high temperature superconductors \cite{R6}, \cite{R7}.          

In the presented paper, we have studied the thermodynamic properties of the anisotropic superconducting state, which can be induced in the metallic molecular hydrogen under the pressure at $414$ GPa (structure $Cmca$). In the present case, three energy gaps can be distinguished, which are connected with the different parts of the Fermi surface. The first and the largest one (symbol $a$) appears on the strongly coupled disk around the $\Gamma$-point; the second gap ($b$) is connected with the "prism-like" sheets; the energy gap of the smallest value ($c$) exists on the remaining regions of the Fermi surface \cite{Cudazzo}. According to the above, the values of the thermodynamic parameters have been calculated with the use of the three-band Eliashberg equations \cite{Eliashberg}: 

\begin{widetext}    
\begin{eqnarray}
\label{r2}
\Delta^\alpha_{n}Z^\alpha_{n}=\pi k_{B}T\sum_{\beta\in\{a,b,c\}}\sum^{M}_{m=-M}
\frac{[K^{\alpha\beta}\left(\omega_{n}-\omega_{m}\right)
-\mu^{\star}_{\alpha\beta}\left(\omega_{m}\right)]}
{\sqrt{\omega^{2}_{m}+\left(\Delta^\beta_{m}\right)^2}} 
{\Delta^\beta_{m}}
\end{eqnarray}
and
\begin{eqnarray}
\label{r3}
Z^\alpha_{n}=1+\pi k_{B}T\sum_{\beta\in\{a,b,c\}}\sum^{M}_{m=-M}\frac{K^{\alpha\beta}
\left(\omega_{n}-\omega_{m}\right)}{\sqrt{\omega^{2}_{m}+
\left(\Delta^\beta_{m}\right)^{2}}}\frac{\omega_{m}}{\omega_{n}} Z^{\beta}_{m}.
\end{eqnarray}
\end{widetext}

The functions $\Delta^\alpha_{n}\equiv\Delta^\alpha\left(i\omega_n\right)$ and  $Z^\alpha_{n}\equiv Z^\alpha\left(i\omega_n\right)$ represent respectively: the order parameters and the wave function renormalization factors; $\omega_n$ is the Matsubara frequency: $\omega_n\equiv\pi k_{B}T\left(2n-1\right)$, where $k_{B}$ is the Boltzmann constant. The symbol $\alpha$ represents the electronic band index. In the considered case $\alpha\in\left\{a, b, c\right\}$.

The functions $K^{\alpha\beta}\left(\omega_n-\omega_m\right)$ define the pairing kernels for the electron-phonon interaction:
\begin{eqnarray}
\label{r4}
K^{\alpha\beta}\left(\omega_n-\omega_m\right)\equiv 2\int^{+\infty}_0 \frac{\alpha^2_{\alpha\beta}F_{\alpha\beta}\left(\Omega\right)\Omega}
{\left(\omega_n-\omega_m\right)^2+\Omega^2}.
\end{eqnarray}
In the presented work the pairing kernels have been approximated in the agreement with the formula below:
\begin{eqnarray}
\label{r5}
K^{\alpha\beta}\left(\omega_n-\omega_m\right)\simeq \lambda^{\alpha\beta}\frac{\Omega^2_C}{\left(\omega_n-\omega_m\right)^2+\Omega^2_C},
\end{eqnarray}
where $\lambda^{\alpha\beta}$ denote the electron-phonon coupling constants ${\lambda^{\alpha\beta}\equiv 2\int^{+\infty}_0 \frac{\alpha^2_{\alpha\beta}\left(\Omega\right) F_{\alpha\beta}\left(\Omega\right)}{\Omega}}$. For the molecular hydrogen under the influence of the pressure at $414$ GPa, the coupling matrix takes the form \cite{Cudazzo}:
\begin{equation}
\label{r6}
\left[\lambda^{\alpha\beta}\right]=\left[\begin{array}{ccc} 0.14 & 0.37 & 1.45\\0.20 & 0.06 & 0.82\\0.11 & 0.11 & 0.71\end{array}\right]. 
\end{equation}

The symbol $\Omega_C$ in Eq. (\ref{r5}) represents the characteristic phonon frequency, which value has been chosen so that the critical temperature estimated on the basis of the Eliashberg equations would be in a fair agreement with the value of $T_{C}$ determined with the help of the {\it ab initio} calculations: $T_{C}=84$ K \cite{Cudazzo}. In particular, in the Eliashberg equations we have assumed that $T=T_{C}$ and then we have decreased the value of the parameter $\Omega_{C}$ until we have reached the equality: $\Delta^{a}_{m=1}=\Delta^{b}_{m=1}=\Delta^{c}_{m=1}=0$. The resulting courses have been presented in Figure 1. We have found that $\Omega_{C}$ equals $107.73$ meV. 

\begin{figure}[ht]
\includegraphics[width=\columnwidth]{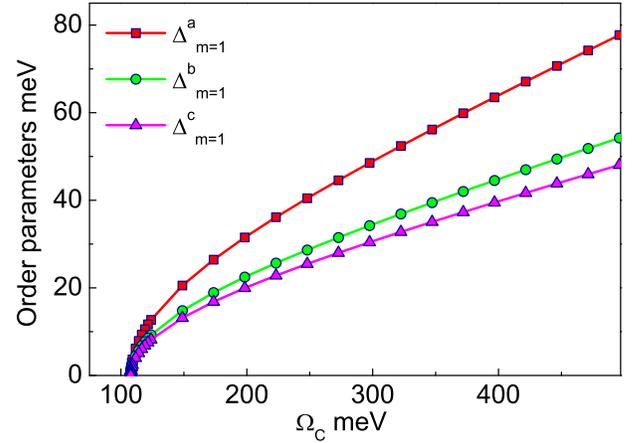}
\label{f1}
\caption{The dependence of the maximum values of the order parameters on $\Omega_{C}$.}
\end{figure}

In the Eliashberg formalism, the Coulomb repulsion between electrons is modeled with the help of the functions: $\mu^{\star}_{\alpha\beta}\left(\omega_m\right)\equiv\mu^{\star}_{\alpha\beta}\theta \left(\omega_{C}-|\omega_m|\right)$,
where $\mu^{\star}_{\alpha\beta}$ are the values of the Coulomb pseudopotentials. The corresponding matrix takes the form \cite{Cudazzo}:
\begin{equation}
\label{r7}
\left[\mu^{\star}_{\alpha\beta}\right]=\left[\begin{array}{ccc} 0.046 & 0.010 & 0.100\\0.006 & 0.046 & 0.180\\0.007 & 0.025 & 0.182\end{array}\right]. 
\end{equation}
\begin{figure*}[ht]
\includegraphics[scale=0.60]{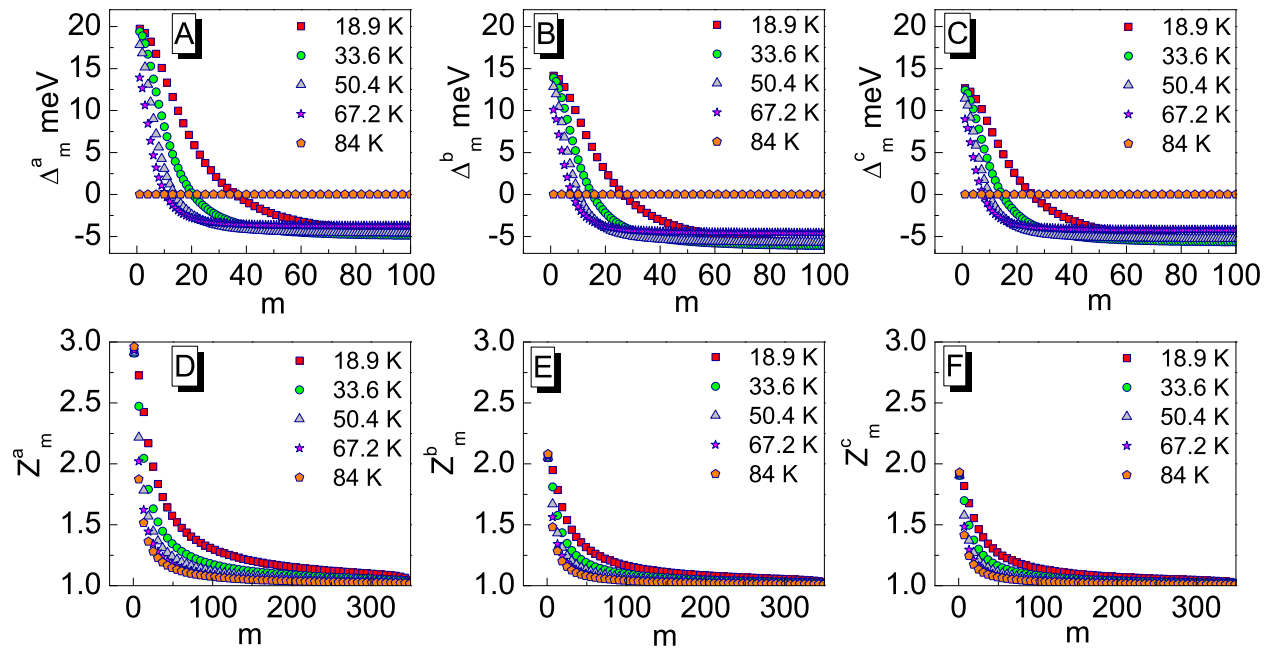}
\label{f2}
\caption{(A)-(C) The values of the order parameters and (D)-(F) the wave function renormalization factors on the imaginary axis.}
\end{figure*}

The quantity $\omega_C$ determines the cut-off frequency: $\omega_{C}=5\Omega_{\rm max}$, where $\Omega_{\rm max}$ denotes the value of the maximum phonon frequency ($\Omega_{\rm max}=496$ meV). 

The Eliashberg equations have been solved for $701$ Matsubara frequency ($M=350$). We have used the numerical methods presented in the works \cite{R8}, \cite{R9}. In the considered case, the functions $\Delta^\alpha_{n}$ and $Z^\alpha_{n}$ are stable for $T\geq T_{0}=18.9$ K.

In Figure 2, we have presented the form of the order parameters and the wave function renormalization factors on the imaginary axis for the selected values of the temperature. It can be easily noticed that the considered functions have the characteristic Lorentzian shape; wherein the order parameters saturate much faster than the wave function renormalization factors.

The temperature dependence of the functions $\Delta^{\alpha}_{m}$ and $Z^{\alpha}_{m}$ can be investigated in the easiest way after
plotting the courses of $\Delta^{\alpha}_{m=1}\left(T\right)$ and  $Z^{\alpha}_{m=1}\left(T\right)$ (Figure 3 (A) and (B)). Let us note that the curves plotted in Fig. 3 (A) can be parametrized with the help of the function $\Delta^{\alpha}_{m=1}\left(T\right)=\Delta^{\alpha}_{m=1}\left(T_{0}\right)\sqrt{1-\left(\frac{T}{T_{C}}\right)^{\eta}}$, where the parameter $\Delta^{\alpha}_{m=1}\left(T_{0}\right)$ takes the values: $19.73$ meV, $14.15$ meV and $12.64$ meV. The exponent $\eta$ is equal to $3.2$. From the physical point of view, the functions $2\Delta^{\alpha}_{m=1}\left(T\right)$ with a good approximation reproduce the
temperature dependence of the energy gaps on the relevant parts of the Fermi surface.

Based on the results presented in Figure 3 (B), it has been stated that the maximum values of the wave function renormalization factors very little depend on the temperature. Let us notice that the functions $Z^{\alpha}_{m=1}\left(T\right)$ reproduce the values of the electron effective mass ($\left[m^{\star}_{e}\right]_{\alpha}$) in a given band with a good approximation. In particular: 
$\left[m^{\star}_{e}\right]_{\alpha}\simeq Z^{\alpha}_{m=1}\left(T\right)\left[m_{e}\right]_{\alpha}$, where the symbol  $\left[m_{e}\right]_{\alpha}$ stands for the electronic band mass. Thus, the obtained results show that the effective mass of the electrons very little depends on the temperature in the whole range of the existence of the superconducting state, and it reaches relatively high values.    

For $T=T_{C}$, the numerical results obtained for the wave function renormalization factors can be
compared with the rigorous analytical results. In particular, the formula below is true: 
\begin{equation}
\label{r8}
Z^{\alpha}_{m=1}\left(T_{C}\right)=1+\sum_{\beta\in\{a,b,c\}}\lambda^{\alpha\beta}.
\end{equation}
The numerical results have the following forms: $Z^{a}_{m=1}\left(T_{C}\right)=2.96$, $Z^{b}_{m=1}\left(T_{C}\right)=2.08$ and $Z^{c}_{m=1}\left(T_{C}\right)=1.93$. Using the expressions \eq{r6} and (8), it can be proved that the analytical approach gives the same results, which demonstrates a high accuracy of the numerical analysis. 

\begin{figure}[ht]
\includegraphics[width=\columnwidth]{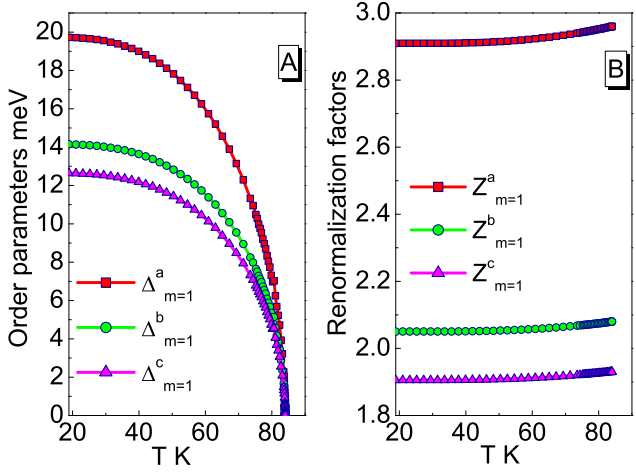}
\label{f3}
\caption{(A) The maximum values of the order parameters and (B) the maximum values of the wave function renormalization factors in the range of the
temperature from $T_{0}$ to $T_{C}$.}
\end{figure}

The exact physical value of the order parameters can be estimated based on the knowledge of the order parameter functions on the real axis ($\Delta^{\alpha}\left(\omega\right)$). Considered functions can be calculated on the basis of the formula \cite{Beach}:
\begin{equation}
\label{r9}
\Delta^{\alpha}\left(\omega\right)=\frac{p^{\alpha}_{1}+p^{\alpha}_{2}\omega+...+p^{\alpha}_{r}\omega^{r-1}}
{q^{\alpha}_{1}+q^{\alpha}_{2}\omega+...+q^{\alpha}_{r}\omega^{r-1}+\omega^{r}},
\end{equation}
where $p^{\alpha}_{j}$ and $q^{\alpha}_{j}$ are the number coefficients; whereas the exponent $r$ is equal to $7$. Then, the equation below should be solved:
\begin{equation}
\label{r10}
\Delta\left(T\right)={\rm Re}\left[\Delta\left(\omega=\Delta\left(T\right),T\right)\right].
\end{equation}

From the physical point of view, the most interesting result has been being obtained for the lowest considered temperature. Figure 4 presents the form of the order parameters for $T=T_{0}$ in the range of the frequency from $0$ to $70$ meV. On that basis, the values of the dimensionless ratios have been determined: $R_{\Delta^{\alpha}}\equiv 2\Delta^{\alpha}\left(0\right)/k_{B}T_{C}$, where 
$\Delta^{\alpha}\left(0\right)\simeq \Delta^{\alpha}\left(T_{0}\right)$. The following results have been obtained: $R_{\Delta^{a}}=5.55$, $R_{\Delta^{b}}=3.96$ and $R_{\Delta^{c}}=3.53$. Referring to the BCS theory \cite{BCS1}, \cite{BCS2}, which predicts that $R_{\Delta}=3.53$, one can see a particularly strong deviation from the BCS result for the band $a$. 

\begin{figure}[ht]
\includegraphics[width=\columnwidth]{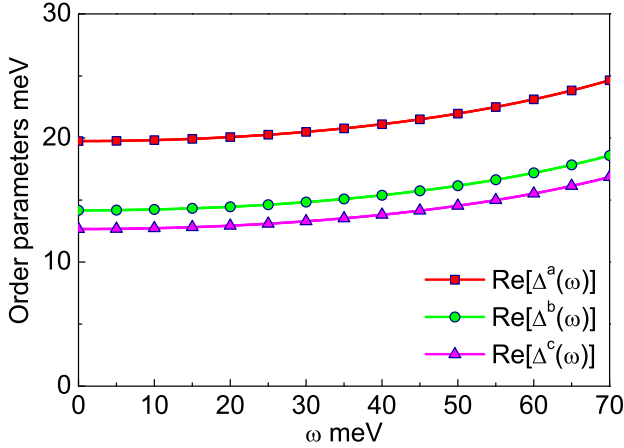}
\label{f4}
\caption{The form of the functions ${\rm Re}\left[\Delta^{\alpha}\left(\omega\right)\right]$ on the real axis. The functions  
${\rm Im}\left[\Delta^{\alpha}\left(\omega\right)\right]$ in the considered area of the frequencies assume the value equal to zero.}
\end{figure}

In the next step, using the results achieved for the order parameter on the real axis, the total normalized density of states has been determined:
\begin{equation}
\label{r11}
\frac{N_{S}\left(\omega\right)}{N_{N}\left(\omega\right)}\equiv\sum_{\alpha\in\{a,b,c\}}\chi_{\alpha}
\frac{N^{\alpha}_{S}\left(\omega\right)}{N^{\alpha}_{N}\left(\omega\right)}, 
\end{equation} 
where the symbols $N^{\alpha}_{S}\left(\omega\right)$ and $N^{\alpha}_{N}\left(\omega\right)$ denote the density functions for the superconducting and normal state in the range of a given band. In addition, the following weights have been estimated: $\chi_{\alpha}=1/3$. The normalized band densities of states have been calculated using the formula:
\begin{equation}
\label{r12}
\frac{N^{\alpha}_{S}\left(\omega\right)}{N^{\alpha}_{N}\left(\omega\right)}={\rm Re}\left[\frac{|\omega-i\Gamma|}
{\sqrt{\left(\omega-i\Gamma\right)^{2}-\left(\Delta^{\alpha}\left(\omega\right)\right)^{2}}}\right], 
\end{equation} 
where the pair breaking parameter $\Gamma$ equals $0.15$ meV. 

In Figure 5, the form of the total normalized density of states for the selected values of the temperature has been plotted. Based on the presented data, it can be noticed that the characteristic peaks of the function $N_{S}\left(\omega\right)/N_{N}\left(\omega\right)$ appear in the points $\omega=\pm\Delta^{\alpha}$.
\begin{figure}[ht]
\includegraphics[width=\columnwidth]{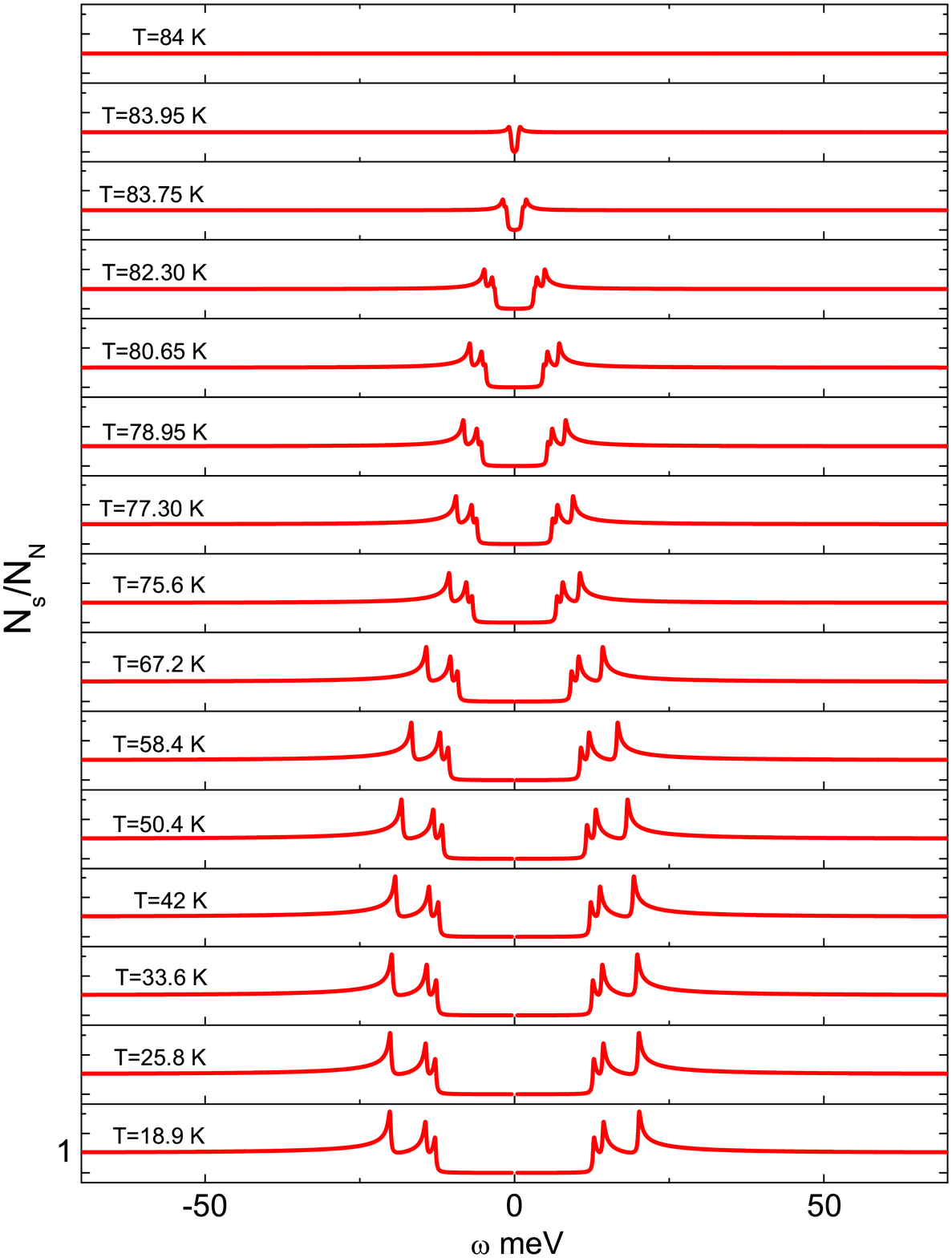}
\label{f5}
\caption{The total normalized density of states in the range of the temperatures from $T_{0}$ to $T_{C}$.}
\end{figure}

After the analytical continuation of the functions $Z^\alpha_{n}$ on the real axis, the exact values of the electron effective mass for a given band can be determined. Up to this point, the following formula should be used: 
$\left[m^{\star}_{e}\right]_{\alpha}={\rm Re}\left[Z^{\alpha}\left(0\right)\right]\left[m_{e}\right]_{\alpha}$. On the basis of the conducted calculations, it has been found that the values of the effective mass are very weakly dependent on the temperature and reach a maximum at a critical temperature. In particular:  $\left[m^{\star}_{e}\right]_{a}=2.99 \left[m_{e}\right]_{a}$, $\left[m^{\star}_{e}\right]_{b}=2.10 \left[m_{e}\right]_{b}$ and $\left[m^{\star}_{e}\right]_{c}=1.94 \left[m_{e}\right]_{c}$. The full forms of the functions $Z^{\alpha}\left(\omega\right)$ for $T=T_{C}$ have been presented in Figure 6.

The thermodynamic critical field and the specific heat should be calculated on the basis of the free energy difference between
the superconducting and normal state:
\begin{widetext}
\begin{equation}
\label{r13}
\Delta F=-2\pi k_{B}T\sum^{M}_{m=1}\sum_{\alpha\in\{a,b,c\}}\rho_\alpha\left(0\right)
[\sqrt{\omega^2_m+\left(\Delta^\alpha_m\right)^2}-|\omega_m|]
[Z^{\alpha,\left(S\right)}_m-Z^{\alpha,\left(N\right)}_m \frac{|\omega_m|}{\sqrt{\omega^2_m+\left(\Delta^\alpha_m\right)^2}}],
\end{equation}
\end{widetext}
where the upper indexes $S$ and $N$ denote the superconducting and normal state, respectively. The vector of the density of states at the Fermi level takes the following form \cite{Cudazzo}:
\begin{equation}
\label{r14}
\left[\rho_\alpha\left(0\right)\right]=\left[\begin{array}{c} 0.067 \\0.126 \\0.908 \end{array}\right] {\rm \frac{states}{eV*cell}}. 
\end{equation}
\begin{figure}[ht]
\includegraphics[width=\columnwidth]{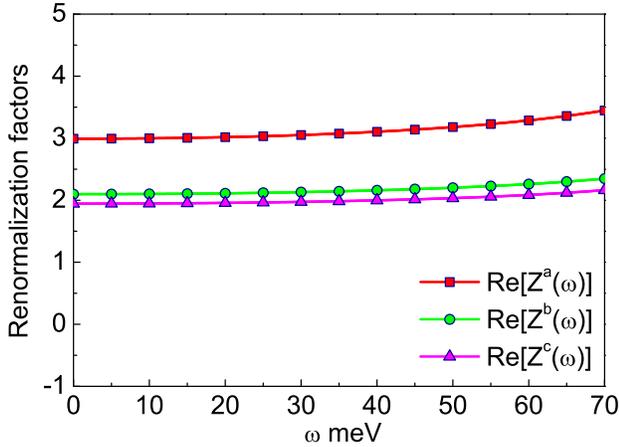}
\label{f6}
\caption{The form of ${\rm Re}\left[Z^{\alpha}\left(\omega\right)\right]$ on the real axis. In the considered area of the
frequencies, the functions ${\rm Im}\left[Z^{\alpha}\left(\omega\right)\right]$ assume the value equal to zero.}
\end{figure}
\begin{figure}[ht]
\includegraphics[width=\columnwidth]{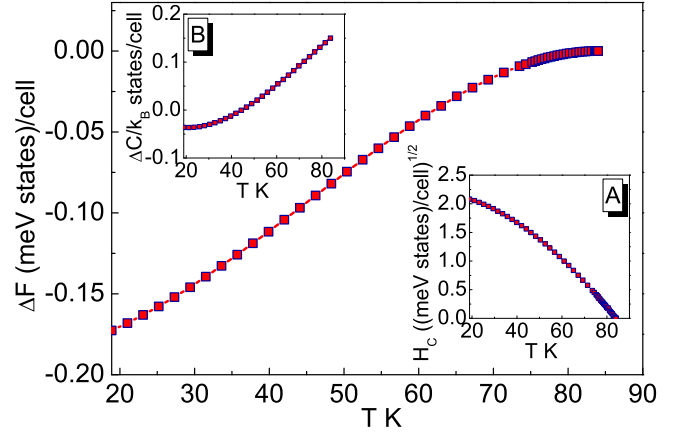}
\label{f7}
\caption{The free energy difference between the superconducting and normal state as a function of the temperature. The inset (A) - the thermodynamic critical field as a function of the temperature. The inset (B) - the specific heat difference between the superconducting and normal state as a function of the temperature.}
\end{figure}

The thermodynamic critical field should be estimated on the basis of the expression below: 
\begin{equation}
\label{r15}
H_C=\sqrt{-8\pi\Delta F}.
\end{equation}

The specific heat difference between the superconducting and normal state can be calculated using the formula: 
\begin{equation}
\label{r16}
\Delta C=-T\frac{d^2\Delta F}{dT^2}.
\end{equation}

In Figure 7, we have provided the plot of the dependence of the free energy difference on the temperature. Additionally, the insets (A) and (B) present the form of the thermodynamic critical field and the specific heat difference in the range of the temperatures from $T_{0}$ to $T_{C}$. Comparing the achieved results to the results obtained in the framework of the one-band model for the comparable value of the pressure ($p=428$ GPa), it has been found that the discussed functions generally reach much larger values. In particular for $T=T_{0}$, the following has been achieved
$\left[\Delta F\right]_{p=414 {\rm GPa}}/\left[\Delta F\right]_{p=428 {\rm GPa}}=6.34$ and
$\left[H_{C}\right]_{p=414 {\rm GPa}}/\left[H_{C}\right]_{p=428 {\rm GPa}}=1.72$; whereas for the critical temperature the results are: 
$\left[\Delta C\right]_{p=414 {\rm GPa}}/\left[\Delta C\right]_{p=428 {\rm GPa}}=4.34$. From the physical point of view, the above results are linked with the relatively high value of the electron density of states in the bands $b$ and $c$.      

Summing up, the basic thermodynamic parameters of the superconducting state in the metallic molecular hydrogen have been provided in the presented work. The pressure value equal to $414$ GPa has been considered.  

It has been stated that the dimensionless ratios of the order parameter to the critical temperature ($2\Delta^{\alpha}\left(0\right)/k_{B}T_{C}$) are equal to: $5.55$, $3.96$ and $3.53$, respectively. It can be easily noticed that the first value significantly exceeds the value predicted by the classical BCS model. Let us draw our attention the fact that the courses of the order parameters on the real axis allowed us to determine the total normalized function of the density of states for the selected values of the temperature.

Additionally, the following quantities have been calculated: the free energy difference, the thermodynamic critical field and the specific heat difference. On the basis of the achieved results, it has been concluded that the values of the considered functions are highly underestimated in the framework of the one-band model.  

In the last step, the wave function renormalization factors have been determined. It ensured the calculation of the maximum values of the electron effective mass in a given band ($\left[m^{\star}_{e}\right]_{\alpha}$). The following results have been
obtained: $\left[m^{\star}_{e}\right]_{a}=2.99 \left[m_{e}\right]_{a}$, $\left[m^{\star}_{e}\right]_{b}=2.10 \left[m_{e}\right]_{b}$ and $\left[m^{\star}_{e}\right]_{c}=1.94 \left[m_{e}\right]_{c}$. 

\begin{acknowledgments}
The authors would like to thank K. Dzili{\'{n}}ski for providing excellent working conditions and financial support.

Some calculations have been conducted on the Cz{\c{e}}stochowa University of Technology cluster, built in the framework of the
PLATON project, no. POIG.02.03.00-00-028/08 - the service of the campus calculations U3. 
\end{acknowledgments}


%
\end{document}